\documentclass[slac_one]{revtex4}
\usepackage{graphicx}
\usepackage{fancyhdr}
\newcommand{\mevcc}{\!\mathrm{MeV}/c^2}

\newcommand{\etac}{\eta_{c}(1S)}
\newcommand{\etacprime}{\eta_{c}(2S)}
\newcommand{\hc}{h_{c}(1P)}
\newcommand{\chicj}{\chi_{cJ}(1P)}
\newcommand{\chiczero}{\chi_{c0}(1P)}
\newcommand{\chicone}{\chi_{c1}(1P)}
\newcommand{\chictwo}{\chi_{c2}(1P)}

\newcommand{\etaprime}{\eta^{\prime}}

\newcommand{\pimp}{\pi^{\mp}}
\newcommand{\pip}{\pi^{+}}
\newcommand{\pim}{\pi^{-}}
\newcommand{\piz}{\pi^{0}}
\newcommand{\kpm}{K^{\pm}}

\newcommand{\kp}{K^{+}}
\newcommand{\km}{K^{-}}
\newcommand{\ks}{K_{S}}
\newcommand{\jpsi}{J/\psi(1S)}
\newcommand{\psip}{\psi(2S)}

\begin{document}

\title{An Experimental Review of Charmonium} 

%

\author{R. E. Mitchell}
\affiliation{Indiana University, Bloomington, IN 47405, USA}

\begin{abstract}
This review briefly outlines recent experimental results from the charmonium system.  These include new measurements of the M1 radiative transition rates $B(\psi(1S,2S)\to\gamma\etac)$; new insights into the mass of the $\etac$; the observation of $\jpsi$ decays to three photons;  new measurements of the two-photon widths of the $\chicj$;  the observation of radiative decays of the $\chicj$ to light quarks;  a precise measurement of the $\hc$ mass;  and an update on properties of the $\etacprime$.  Each result adds piecemeal to our understanding of the strong force.
\end{abstract}

\maketitle

\thispagestyle{fancy}


\section{INTRODUCTION} 

After more than 30 years, the charmonium system continues to provide an important laboratory for the study of the strong force.  This brief review reflects the wide scope of strong force physics that is accessible in the charmonium system.  The physics impact of the results presented here ranges from tests of lattice QCD calculations to our understanding of bound gluonic states (glueballs) to an evaluation of models of the spin-spin component of the $q\overline{q}$ potential to our ability to calculate strong radiative and relativistic corrections.   In small increments, the charmonium system has, over time, given us a variety of opportunities to add to our global understanding of the strong force\footnote{For a comprehensive review of the range of theoretical ideas experimentally accessible in the charmonium system, refer to~\cite{QWGYellow} or~\cite{BESPhysics}; also refer to the references cited in the experimental papers described below.}.

The charmonium system can be accessed experimentally in a number of complementary ways.  This review covers recent results obtained by three different methods.  (1)~The CLEO-c experiment at Cornell University collects data at an $e^{+}e^{-}$ collider (CESR) that has a center of mass energy in the charmonium region.  CLEO-c has recently collected a total and final sample of 27~million $\psip$ decays.  Many other charmonium states can be reached and studied through transitions of the $\psip$.  (2,3)~The Belle (at KEKB) and BaBar (at SLAC) experiments run at $e^{+}e^{-}$ colliders with center of mass energies in the bottomonium region.  At these higher energies, among other possibilities, one can (2)~study the collisions of two photons radiated from the $e^{+}e^{-}$, where the center of mass energy of the two photons is in the charmonium region; or one can (3)~study charmonium states produced in $B$ decays.  

Results from BES-II are not covered in this review.  It should be noted, however, that BES-III is on the horizon and will soon increase, substantially, the world sample of $\jpsi$ and $\psip$ decays.

\section{THE M1 TRANSITIONS {\Large\boldmath $\psi(1S,2S)\to\gamma\etac$} }

\begin{figure*}[t]
\centering
\includegraphics*[width=3.03in]{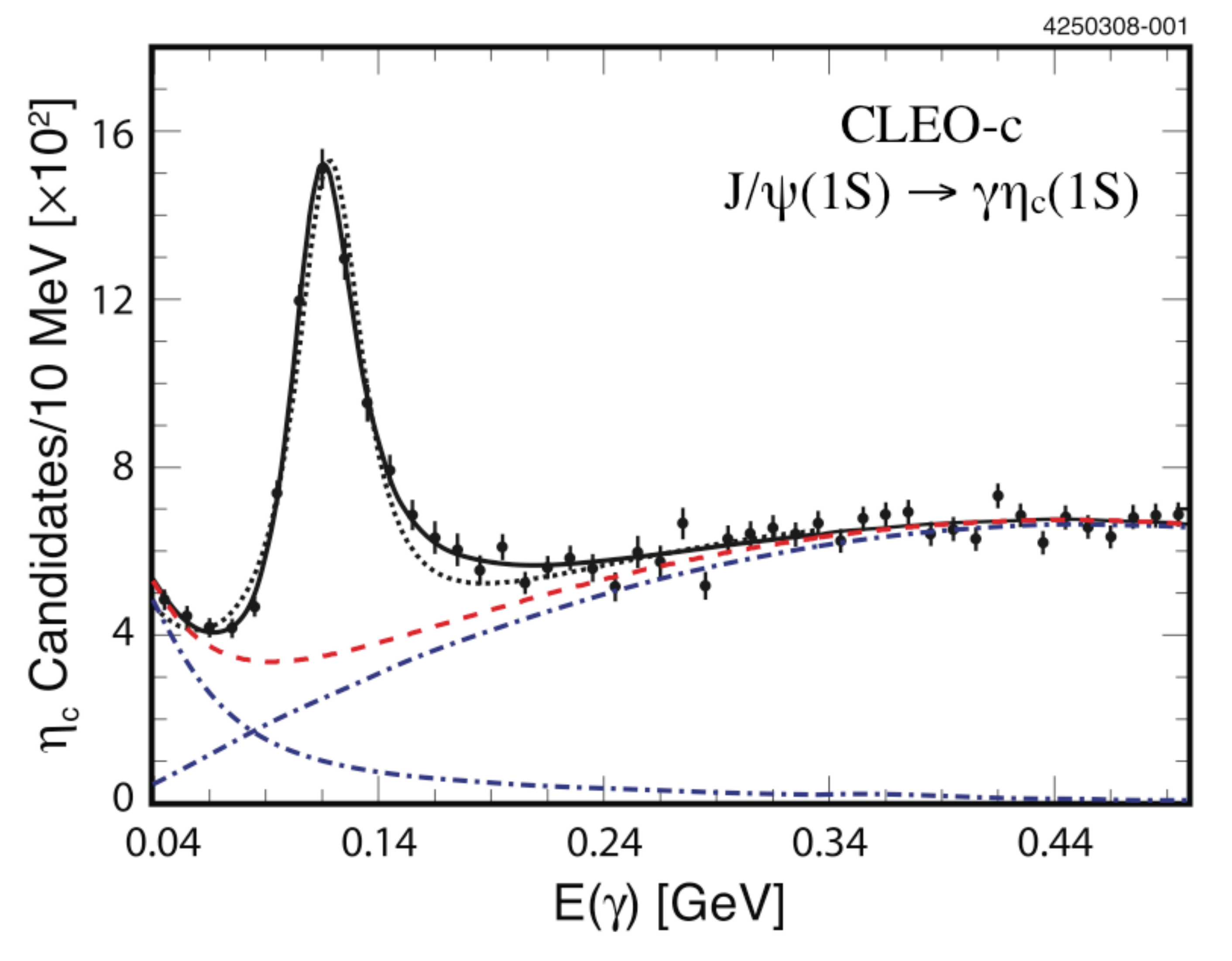}
\includegraphics*[width=3.58in]{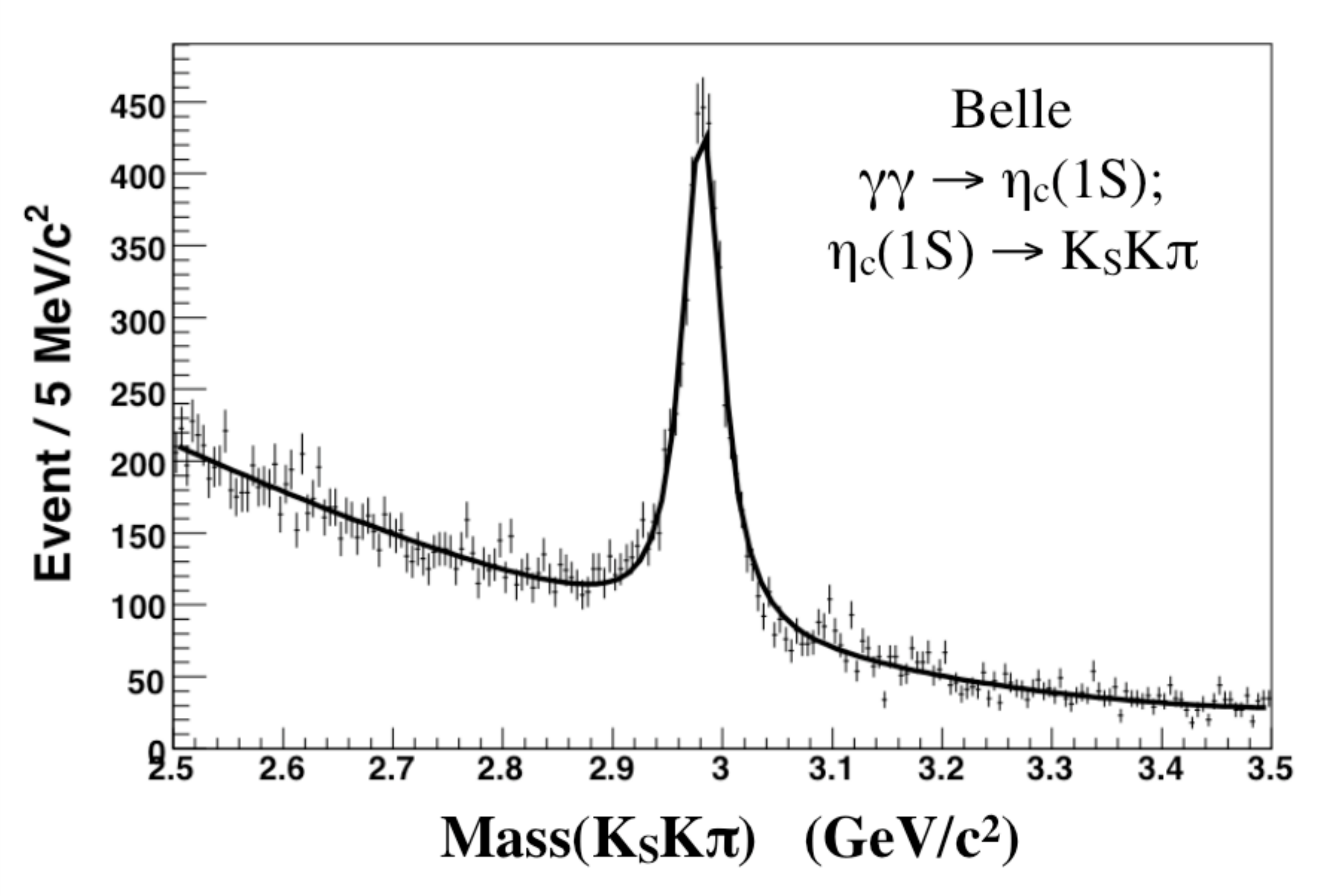}
\caption{(Left) The photon energy spectrum from $\jpsi\to\gamma\etac$ from the CLEO-c experiment~\cite{JPsiGammaEtac}.  The $\etac$ is reconstructed in 12 different exclusive decay modes.  The dotted line is a fit with a relativistic Breit-Wigner; the solid line uses a relativistic Breit Wigner distorted by the energy dependences of the phase space term and the matrix element.  (Right) The distribution of Mass($\ks\kpm\pimp$) obtained from the process $\gamma\gamma\to\ks\kpm\pimp$ from the Belle experiment~\cite{BELLEKKpi}.} \label{fig:etac}
\end{figure*}

Radiative transitions in the charmonium system have recently been the subject of both lattice QCD calculations~\cite{m1lqcd} and effective field theory techniques~\cite{m1eft}.  Key among these are the magnetic dipole (M1) transitions $\jpsi\to\gamma\etac$ and $\psip\to\gamma\etac$.  Using a combination of inclusive and exclusive techniques, the CLEO-c experiment has recently measured $B(\jpsi\to\gamma\etac) = (1.98\pm0.09\pm0.30)\%$ and $B(\psip\to\gamma\etac) = (4.32\pm0.16\pm0.60)\times10^{-3}$~\cite{JPsiGammaEtac}, reconciling experiment with theory.  The line shape of the $\etac$ in these M1 transitions was found to play a crucial role.  Because the width of the $\etac$ is relatively large, the energy dependences of the phase space term and the matrix element distort the line shape (see Figure~\ref{fig:etac}a).  The theoretical uncertainty in this line shape represents the largest systematic error in the branching ratios.  Both M1 transitions are larger than previous measurements due to a combination of a larger $\etac$ width and an accounting for the asymmetry in the line shape.

\section{MASS OF THE {\Large\boldmath $\etac$} }

Current lattice QCD calculations of the $1S$ hyperfine splitting ($M(\jpsi)-M(\etac)$) have MeV-level sensitivity.  The mass of the $\etac$ carries the dominant experimental error in checking these calculations.  As of 2006, there was a $3.3~\!\sigma$ discrepancy between $\etac$ mass measurements made from $\psi(1S,2S)\to\gamma\etac$ (averaging $2977.3\pm1.3~\mevcc$) and from $\gamma\gamma$ or $p\overline{p}$ production (averaging $2982.6\pm1.0~\mevcc$)~\cite{PDG2006}.  The CLEO-c analysis referenced above~\cite{JPsiGammaEtac} suggests that the solution to this problem may lie in the line shape of the $\etac$ in the M1 radiative transitions (see Figure~\ref{fig:etac}a).  When no distortion in the line shape is used, the resulting $\etac$ mass is consistent with other measurements from M1 transitions ($2976.6\pm0.6~\mevcc$, statistical error only); however, when a distorted line shape is taken into account, the mass is consistent with those from $\gamma\gamma$ or $p\overline{p}$ production ($2982.2\pm0.6~\mevcc$, statistical error only).  Recent measurements of the $\etac$ mass in $\gamma\gamma$ production by the Belle experiment are consistent with this general picture.  In an analysis of $\gamma\gamma\to\ks\kpm\pimp$, Belle measures a mass of $2981.4\pm0.5\pm0.4~\mevcc$ (see Figure~\ref{fig:etac}b)~\cite{BELLEKKpi}\footnote{Note that when interference with the continuum is included, the $\etac$ mass is slightly higher.}.  An $\etac$ mass of $2986.1\pm1.0\pm2.5~\mevcc$ is reported in a Belle analysis of $\gamma\gamma\to h^{+}h^{-}h^{+}h^{-}$, where $h = \pi,K$~\cite{BELLE4Body}.  Both measurements are consistent with the higher $\etac$ mass.

\section{OBSERVATION OF {\Large\boldmath $\jpsi\to\gamma\gamma\gamma$} }

The $^{3}S_{1}$ state of positronium, ortho-positronium, decays almost exclusively to $3\gamma$ and has played an important role in precision tests of QED.  The analogous decay in the quarkonium system, $\jpsi\to\gamma\gamma\gamma$, has recently been reported by the CLEO-c experiment~\cite{CLEO3Gamma}.  Like ortho-positronium for QED, this $\jpsi$ decay furthers our understanding of QCD, especially when compared to decays like $\jpsi\to\gamma gg$, $\jpsi\to ggg$, or $\jpsi\to l^{+}l^{-}$ due to similarities at the parton level~\cite{jpsi3gamma}.  Excluding backgrounds from $\jpsi\to\gamma(\eta,\piz,\etaprime)\to\gamma\gamma\gamma$ and carefully accounting for backgrounds with missing photons (e.g., $\jpsi\to\gamma f; f\to\piz\piz$), CLEO-c measures $B(\jpsi\to\gamma\gamma\gamma) = (1.2\pm0.3\pm0.2)\times10^{-5}$ with a significance greater than $6~\!\sigma$.  As part of the same analysis, a 90\% C.L. upper limit of $6\times10^{-6}$ is placed on the product branching fraction $B(\jpsi\to\gamma\etac)\times B(\etac\to\gamma\gamma)$.

\section{TWO PHOTON WIDTHS OF THE {\Large\boldmath $\chicj$} STATES}

Like the decay $\jpsi\to\gamma\gamma\gamma$ discussed above, decays of the $\chicj$ to two photons provide additional analogies to the positronium system.  Masses and wavefunctions cancel in the ratio of $J=2$ to $J=0$ two-photon widths.  Thus, ${\cal R} = \Gamma(^{3}P_{2}\to\gamma\gamma)/\Gamma(^{3}P_{0}\to\gamma\gamma) = 4/15 \approx 0.27$ to leading order for both positronium and charmonium.  Since the lowest order prediction is well-understood, experimental measurements of ${\cal R}$ probe strong radiative and relativistic corrections in the charmonium system.  

\begin{figure*}[t]
\centering
\includegraphics*[width=3.58in]{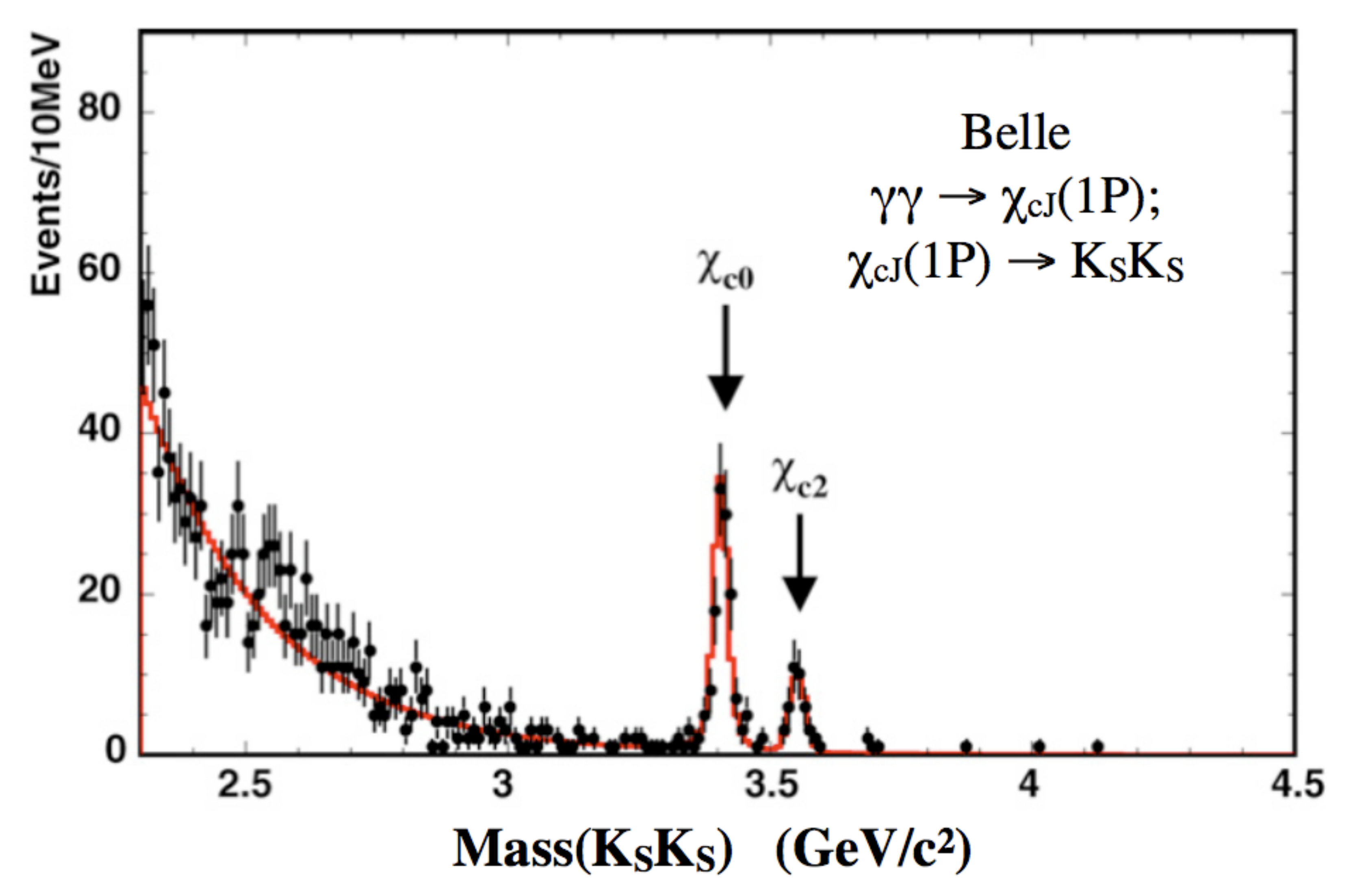}
\includegraphics*[width=3.2in]{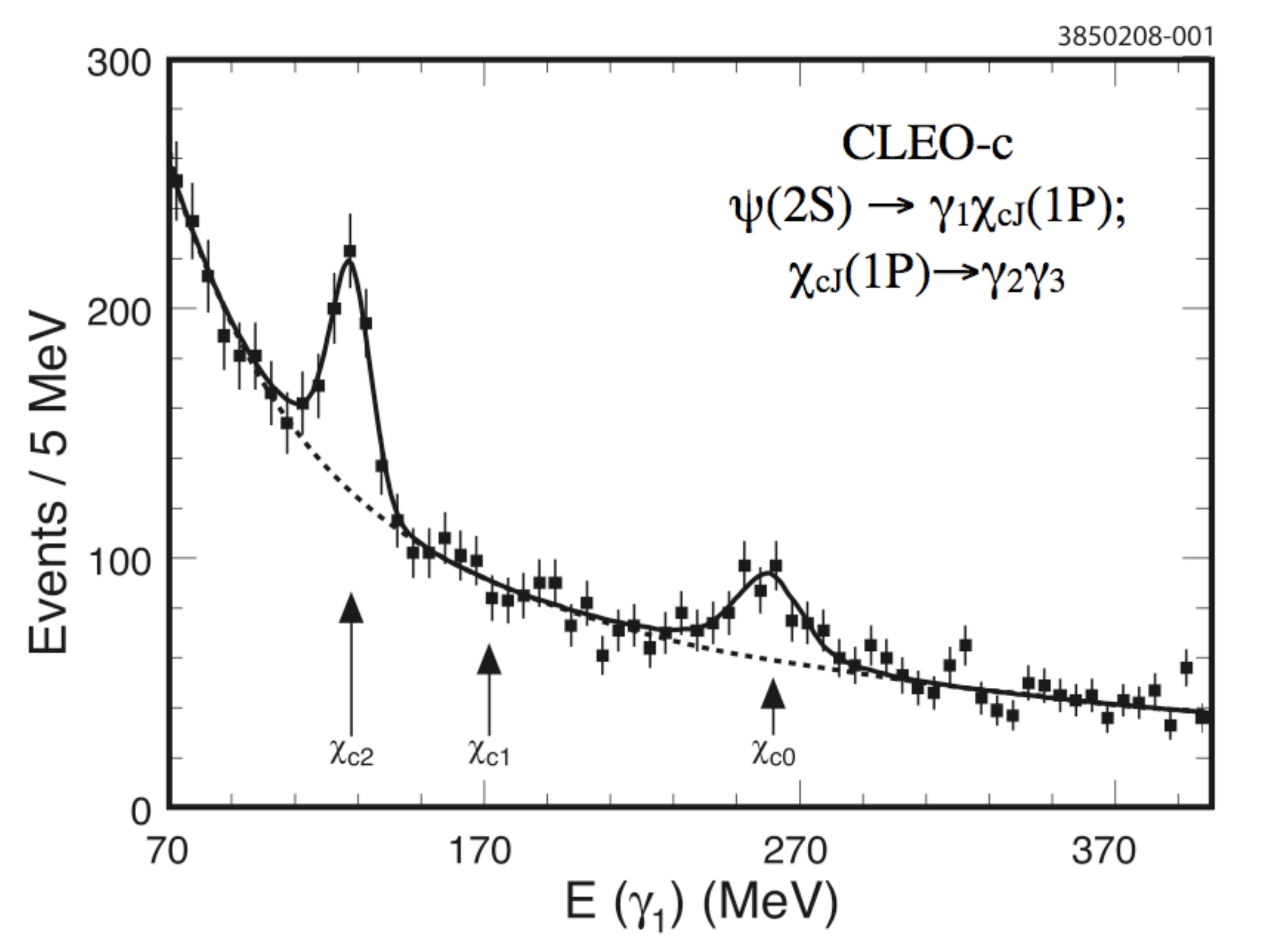}
\caption{(Left) The distribution of Mass($\ks\ks$) obtained from the process $\gamma\gamma\to\ks\ks$ from the Belle experiment~\cite{BELLE2Body}.  Dividing by an external value of $B(\chicj\to\ks\ks)$, one can obtain the two-photon widths of the $\chicj$ in a ``formation'' process.  (Right) The photon energy spectrum of the lowest energy photon from $\psip\to\gamma\chicj; \chicj\to\gamma\gamma$ from the CLEO-c experiment~\cite{CLEO2Gamma}.  Dividing by external values of $B(\psip\to\gamma\chicj$) and the total widths of the $\chicj$, the two-photon widths of the $\chicj$ can be extracted in a ``decay'' process.} \label{fig:chic}
\end{figure*}

Belle has recently measured the product of $\chicj$ branching fractions times two-photon widths of the $\chiczero$ and $\chictwo$ in the ``formation'' process $\gamma\gamma\to\chicj; \chicj\to X$, where $X$ is a hadronic decay mode of the $\chicj$.  The two-photon widths are obtained by dividing by external determinations of $B(\chicj\to X)$.  When $X = \ks\ks$  (see Figure~\ref{fig:chic}a), the Belle value is ${\cal R} = 0.18\pm0.03\pm0.04$~\cite{BELLE2Body}. When $X = \pip\pim\pip\pim, \kp\km\pip\pim, \kp\km\kp\km$, the Belle values are ${\cal R} = 0.22\pm0.03\pm0.05, 0.21\pm0.03\pm0.09, 0.21\pm0.05\pm0.06$, respectively~\cite{BELLE4Body}.

CLEO-c has recently measured the same quantity in the ``decay'' process $\psip\to\gamma\chicj; \chicj\to\gamma\gamma$ (see Figure~\ref{fig:chic}b), obtaining two-photon widths after dividing by external determinations of $B(\psip\to\gamma\chicj)$ and the total widths of the $\chicj$~\cite{CLEO2Gamma}. The ratio in this case is ${\cal R} = 0.237\pm0.043_{stat}\pm0.015_{syst}\pm0.03_{PDG}$, consistent with the determination from the ``formation'' processes.  The world-average value of ${\cal R}$, including the values presented here, is now $0.20\pm0.02$.  

\section{OBSERVATION OF {\Large\boldmath $\chicone\to\gamma(\rho,\omega)$} }

Radiative decays of the $\jpsi$ have often been used as a gluon-rich environment in which one can search for bound states of gluons (glueballs).  The successful theoretical interpretation of these searches, however, relies on an understanding of the decays $\jpsi\to\gamma f_{0}$, where the $f_{0}$ states are complicated by the fact that they are likely mixtures of conventional states and a glueball.  Structurally similar to these $\jpsi$ decays, the decays $\chicj\to\gamma(\rho,\omega,\phi)$ do not suffer the same complications since the quark model interpretations of the $\rho$, $\omega$, and $\phi$ mesons are well-established.  Therefore, these $\chicj$ radiative decays provide an independent and complementary $c\overline{c}$-annihilation decay that can be used to validate theoretical techniques.  CLEO-c has recently observed the decays $\chicone\to\gamma\rho$ and $\chicone\to\gamma\omega$ with branching fractions of $(2.43\pm0.19\pm0.22)\times10^{-4}$ and $(8.3\pm1.5\pm1.2)\times10^{-5}$, respectively~\cite{ChicRad}.  Pertubative QCD (pQCD) predictions, however, are more than an order of magnitude smaller~\cite{ChicRadTheory}.  A resolution to this anomaly has yet to be found.

\section{MASS OF THE {\Large\boldmath $\hc$} }

Because the masses of the $\chicj$ are already well-measured, improving the mass measurement of the $\hc$ yields important information regarding the hyperfine splitting of the $1P$~states.  The hyperfine splitting of the $1P$~states is in turn sensitive to the form of the $q\overline{q}$ potential.  In particular, for the $1P$ states, if there is no spin-spin contribution from the confinement part of the potential, it is expected that the hyperfine splitting will be zero to lowest order.  CLEO-c has recently made a precision measurement of the mass of the $\hc$ in the processes $\psip\to\piz\hc; \hc\to\gamma\etac$, where the $\etac$ decay is either reconstructed inclusively or through a series of exclusive modes~\cite{CLEOHc}.  In the inclusive analysis, CLEO-c measures $M(\hc) = 3525.35\pm0.23\pm0.15~\mevcc$, while in the exclusive analysis $M(\hc) = 3525.21\pm0.27\pm0.14~\mevcc$.  Combining these results, CLEO-c obtains $M(\hc) = 3525.28\pm0.19\pm0.12~\mevcc$.  To determine the hyperfine splitting, one can compare this result with the spin-weighted centroid of the $\chicj$ masses, $\left < M(\chicj) \right > = 3525.30\pm0.04~\mevcc$.  The difference is $0.02\pm0.19\pm0.13~\mevcc$, thus the hyperfine splitting of the $1P$ states is consistent with zero\footnote{Note that there is some subtlety in the way one calculates the spin-weighted centroid of the $\chicj$ masses;  see the discussion in~\cite{CLEOHc}.}.

\begin{figure*}[t]
\centering
\includegraphics*[width=2.8in]{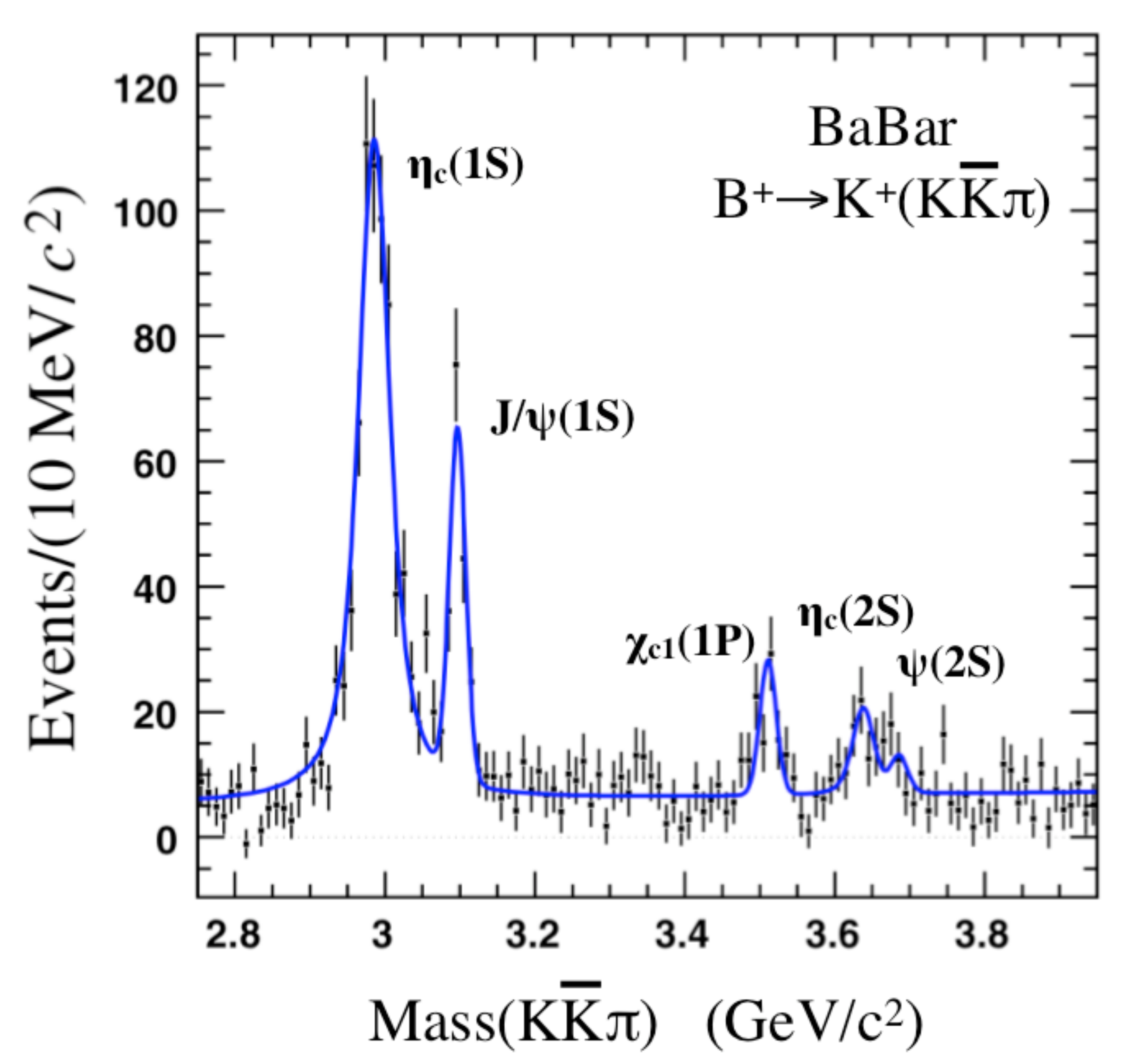}
\includegraphics*[width=3.8in]{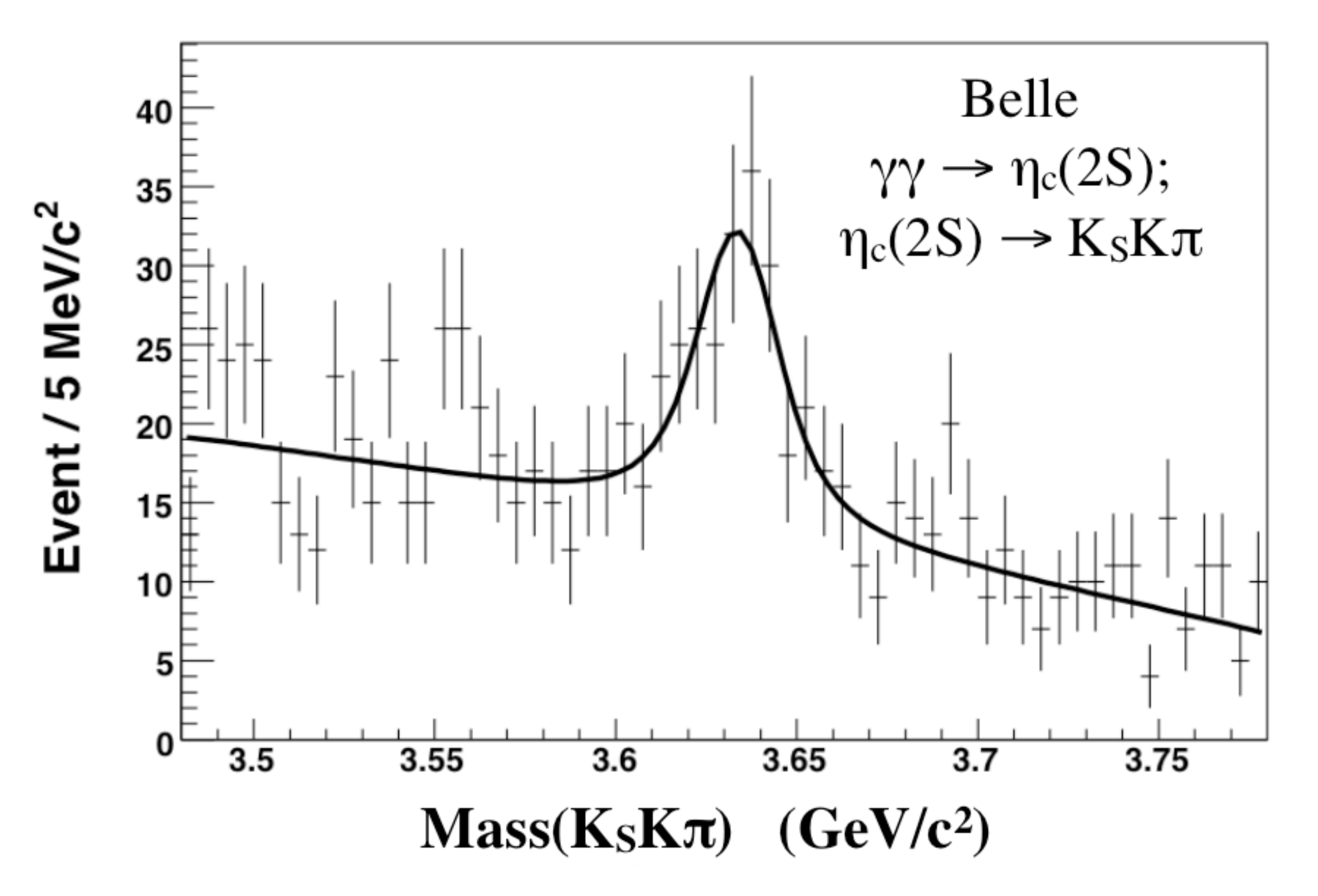}
\caption{(Left) The distribution of Mass($K\overline{K}\pi$) obtained from the process $B^{+}\to\kp(K\overline{K}\pi)$ from the BaBar experiment~\cite{BABAREtacprime}.  Combined with an inclusive measurement, this is used to derive an absolute measurement of $B(\etacprime\to K\overline{K}\pi)$.  (Right)~The distribution of Mass($\ks\kpm\pimp$) showing the $\etacprime$ region obtained from the process $\gamma\gamma\to\ks\kpm\pimp$ from the Belle experiment~\cite{BELLEKKpi}.} \label{fig:etacprime}
\end{figure*}

\section{PROPERTIES OF THE {\Large\boldmath $\etacprime$} }

The $\etacprime$, the first radial excitation of the $\etac$ remains fairly elusive.  Progress has been made, however.  At BaBar, the $\etacprime$ was observed in $B$ decays in its decay mode to $K\overline{K}\pi$~\cite{BABAREtacprime} (see Figure~\ref{fig:etacprime}a).  Combining this result with an earlier inclusive result, one can determine the first absolute branching fraction of the $\etacprime$, $B(\etacprime\to K\overline{K}\pi) = (1.9\pm0.4\pm1.1)\%$.  This is significantly smaller than the corresponding decay of the $\etac$, $B(\etac\to K\overline{K}\pi) = (7.0\pm1.2)\%$.  If one could measure the rate of the process $\gamma\gamma\to\etacprime; \etacprime\to K\overline{K}\pi$ one could now infer the two-photon width of the $\etacprime$, a quantity that is theoretically clean.  Unfortunately, the two-photon process is complicated by interference effects between the $\etacprime$ and the continuum.  Belle, however, can use this process to determine the mass and width of the $\etacprime$ (see Figure~\ref{fig:etacprime}b) as $M(\etacprime) = 3633.7\pm2.3\pm1.9~\mevcc$ and $\Gamma(\etacprime) = 19.1\pm16.9\pm6.0~\mevcc$~\cite{BELLEKKpi}\footnote{The mass and width of the $\etacprime$ change when interference effects are included, but the values are consistent within errors.}.  The decay $\etacprime\to K\overline{K}\pi$ remains the only observed decay mode of the $\etacprime$.

\section{OUTLOOK}

While the CLEO-c and BaBar programs have ended, Belle continues.  In addition, the BES-II experiment at the BEPC $e^{+}e^{-}$ collider has recently been upgraded to BES-III.  The BEPC collider will provide an unprecedented luminosity at center of mass energies in the charmonium region, and we expect BES-III, carrying on in the tradition of BES-II and CLEO-c, to greatly expand our knowledge of the charmonium system.





\end{document}